\documentclass[aps, prd,10pt,notitlepage,nofootinbib,superscriptaddress,showkeys,twocolumn]{revtex4-1}
\usepackage{amstext,amsmath,amssymb,amsfonts,bbm, amsthm}
\usepackage[latin1]{inputenc}
\usepackage{fancyhdr}
\usepackage{graphicx}
\usepackage{hyperref}
\usepackage{tocvsec2}

\usepackage{color}

\def\beq{\begin{equation}}
\def\be{\begin{equation}}
\def\ee{\end{equation}}
\def\bes{\begin{eqnarray}}
\def\ees{\end{eqnarray}}

\def\f{\frac}

\begin{document}

\title{\large \bf Metadiffusion}

\author{{Matteo Smerlak}}\email{smerlak@aei.mpg.de}
\affiliation{Max-Planck-Institut f\"ur Gravitationsphysik, Am M\"uhlenberg 1, D-14476 Golm, Germany}

\date{\small\today}

\begin{abstract}\noindent
We predict that the speed of a diffusion front is modified by tidal forces. This effect is small in astrophysical situations, but could be significant -- indeed tunably so -- in ``gravitational analogues'' such as gradient-index lenses. Confirming this prediction would extend the concept of \emph{metamaterials} to dissipative transport phenomena. 
\end{abstract}
\maketitle

The concept of \emph{metamaterials} \cite{ENGHETA2006} has attracted considerable interest recently. In a nutshell, the idea is to engineer periodic structures in such a way to \emph{tailor} the propagation of waves (electromagnetic or else) within the resulting material. This method is so powerful that it allows to build actual \emph{cloaking devices} \cite{Service2010}, in which light rays incident on an object simply \emph{bypass} it, in effect making it ``invisible''.

The best way to describe this phenomenon theoretically is by analogy with Einstein's theory of general relativity. Recall indeed his famous prediction of 1915: light rays bend when they approach a massive object such as the Sun. This is an effect of the gravitational field, and can be described geometrically in terms of \emph{spacetime curvature}. Similarly, metamaterials can be described as \emph{effective spacetimes}, and the bending of wave they generate as a manifestation of \emph{effective gravitational fields} \cite{Leonhardt}. 

Now, plain propagation is not the only way waves may transport their energy. In highly scattering media, instead, they \emph{diffuse} -- much like sugar diffuses in coffee. This kind of transport, which also arises in the phenomenon of \emph{Brownian motion}, is fundamentally different from the normal, ray-like propagation of waves:  it is \emph{irreversible}. In these so-called \emph{dissipative processes}, ordered patterns (the lamp of sugar) lose their structure with time (the sugar gets dissolved in the coffee) as a diffusion front moves away from the pattern. This dissipation of structure and order into randomness can be quantified by the \emph{diffusion law}: the radius of the diffusion front grows like the \emph{square root of time}. This law follows easily from the \emph{diffusion equation} which governs this dissipative process.

In our recent work on Brownian motion in curved spacetimes \cite{Smerlak2011b}, we established the general form of the diffusion equation in \emph{curved spacetimes}. Thanks to this result, we could investigate whether the diffusion law is modified by the presence of a gravitational field. We found that it is: depending on the kind of tidal forces that the gravitational field generates, a diffusion front will either grow \emph{faster} or \emph{slower} than the square root of time. Due to the small value of Newton's gravitational constant, this effect turned out to be very weak in typical astrophysical situations; it could however probably be measured in one of the gravitational analogues of the kind mentioned earlier, where strong effective gravitational fields can easily be emulated.

\begin{figure}[h!]
\includegraphics[scale=0.45]{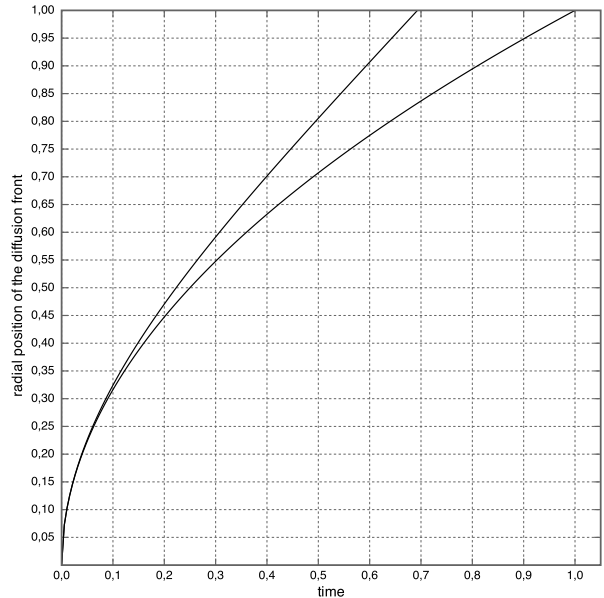}
\caption{A spherical diffusion front in the Maxwell fisheye will grow in time like $R\sqrt{e^{t/R^{2}}-1}$ (top curve) instead of $\sqrt{t}$ (bottom curve); in particular, it will reach the boundary of the lens (top horizontal line) $30\%$ earlier than expected. The units are $R$ for distance and $T=R^{2}/\kappa$ for time, with $\kappa$ the diffusion constant.}
\end{figure}

As an example, we considered the effective spacetime known in optics as the \emph{Maxwell fisheye} \cite{Niven}: an optical ball where the refractive index varies as a function of the radius according to a specific distribution.\footnote{This distribution is $n(r)=\f{n_{0}}{1+(r/R)^{2}}$, where $n_{0}$ is a constant, $r$ the radial coordinate and $R$ the radius of the lens.} We found that, if a pulse of light (an optical relative of a lamp of sugar) diffused from the center of the fisheye, it would reach its boundary approximately $30\%$ earlier than the square-root law would let us expect. Fig. 1 shows the predicted radial position $r$ of a diffusion front in the Maxwell fisheye with radius $R$ versus time, (in units of the diffusion constant).

Furthermore -- and this is the most interesting aspect of our work -- we found that the instantaneous velocity of the diffusion front depends sensitively on the index distribution \cite{Smerlak}. But by designing the lens with a different index profile, or -- even better -- with a time-dependent profile, we could control precisely the motion of the diffusion front: e.g. make it speed up, then slow down, then speed up again -- at will. In other words, it is in principle possible to \emph{tailor diffusion processes} with tunable effective gravitational fields. To our knowledge, this proposal is the first application of the concept of \emph{metamaterials}, in the extended sense of artificial materials engineered to have desired properties, to a \emph{dissipative} transport phenomenon.

The possible applications of this new concept of \emph{metadiffusion} are virtually as many as diffusion phenomena themselves: light diffusion in scattering media, thermal and electric conduction, particle diffusion, osmosis, etc. If we are right, all of these could be tailored by means of effective gravitational fields driven by external parameters.

\bibliographystyle{utcaps}
\providecommand{\href}[2]{#2}\begingroup\raggedright\endgroup
\end{document}